\documentclass{article}

    \PassOptionsToPackage{numbers, compress}{natbib}
 \usepackage[preprint]{neurips_2026}

\usepackage[utf8]{inputenc} 
\usepackage[T1]{fontenc}    
\usepackage{hyperref}       
\usepackage{url}            
\usepackage{booktabs}       
\usepackage{amsfonts}       
\usepackage{nicefrac}       
\usepackage{microtype}      
\usepackage{xcolor}         
\usepackage{xspace}
\usepackage{todonotes}
\usepackage{tikz}
\usepackage{enumitem}
\usepackage{amsmath}
\usepackage{algorithm}
\usepackage{amsmath}
\usepackage{amssymb}
\usepackage{mathtools}
\usepackage{amsthm}
\usepackage{listings}
\usepackage{algorithm}
\usepackage{algorithmic}
\usepackage{subcaption}
\usepackage{amsmath}
\usepackage{amsfonts}
\usepackage{xspace}
\usepackage{xcolor}
\usepackage{multirow}
\usepackage{array}
\usepackage{enumitem}
\usepackage[table]{xcolor}
\definecolor{lightgray}{RGB}{230, 230, 230}
\usepackage{xurl}
\usepackage{wrapfig}
\usepackage{subcaption}
\usepackage{pgfplots}
\usepgfplotslibrary{groupplots}
\usepackage{adjustbox}
\usepackage{wrapfig}

\newenvironment{msenum}
  {\begin{enumerate}[label={\bf \Roman*.}, leftmargin=2em]}
  {\end{enumerate}}

\newcommand{\paragraphB}[1]{\vskip 0pt \noindent {\textbf{#1}.}\xspace}

\usepackage{xspace}
\newcommand{\system}{{{FreeMOCA}}\xspace}

\usepackage[most]{tcolorbox}
\usepackage{xparse}

\usepackage{enumitem}
\newenvironment{smitemize}{
  \begin{itemize}[topsep=1pt, partopsep=2pt, itemsep=3pt, parsep=0pt, leftmargin=10pt, itemindent=1pt]
}{\end{itemize}}

\usetikzlibrary{
    arrows.meta,
    shapes.geometric,
    positioning, 
    calc
}

\definecolor{panelablue}{RGB}{70,130,180}
\definecolor{panelbgreen}{RGB}{76,145,104}
\definecolor{panelcorange}{RGB}{214,132,66}
\definecolor{badred}{RGB}{190,70,70}

\title{\system: Memory-Free Continual Learning for Malicious Code Analysis}

\author{%
  Zahra Asadi\thanks{Joint First Authors} \\
  Department of Computer Engineering\\
  Amirkabir University of Technology\\
  \texttt{zahraasadi257@aut.ac.ir}\\
  \And
  Haeseung Jeon\footnotemark[1]  \\
  Division of Artificial Intelligence \& Software \\
  Ewha Womans University \\
  \texttt{haeseungjeon@ewha.ac.kr} \\
  \And
  Sohyun Han \\
  Division of Artificial Intelligence \& Software \\
  Ewha Womans University \\
  \texttt{hansh329@ewha.ac.kr} \\
  \And
  Md Mahmuduzzaman Kamol \\
  Department of Computer Science \\
  University of Texas at El Paso \\
  \texttt{mkamol@miners.utep.edu} \\
  \And
  Se Eun Oh\thanks{Corresponding Author} \\
  Division of Artificial Intelligence \& Software \\
  Ewha Womans University \\
  \texttt{seoh@ewha.ac.kr} \\
  \And
  Mohammad Saidur Rahman \\
  Department of Computer Science \\
  University of Texas at El Paso \\
  \texttt{msrahman3@utep.edu} \\
}

\begin{document}

\maketitle

\begin{abstract}

As over 200 million new malware samples are identified each year, antivirus systems must continuously adapt to the evolving threat landscape. However, retraining solely on new samples leads to catastrophic forgetting and exploitable blind spots, while retraining on the entire dataset incurs substantial computational cost. We propose FreeMOCA, a memory- and compute-efficient continual learning framework for malicious code analysis that preserves prior knowledge via adaptive layer-wise interpolation between consecutive task updates, leveraging the fact that warm-started task optima are connected by low-loss paths in parameter space. 

We evaluate FreeMOCA in both class-incremental (Class-IL) and domain-incremental (Domain-IL) settings on large-scale Windows (EMBER) and Android (AZ) malware benchmarks. FreeMOCA achieves substantial gains in Class-IL, outperforming 11 baselines on both EMBER and AZ benchmarks.
It also significantly reduces forgetting, achieving the best retention across baselines, and improving accuracy by up to 42\% and 37\% on EMBER and AZ, respectively. These results demonstrate that warm-started interpolation in parameter space provides a scalable and effective alternative to replay for continual malware detection. 

Code is available at: \url{https://github.com/IQSeC-Lab/FreeMOCA}.

\end{abstract}
\section{Introduction}
\label{sec:intro}

Machine learning (ML) has become a standard tool for malware analysis~\cite{malwareguard,haque2025lamda,arp2014drebin,haque2025citadel,sabbah2026understanding,chow2026beyond}. In deployment, however, malware classifiers face {\em non-stationary and highly imbalanced data distributions}, led by adversarial adaptation and the rapid evolution of malware ecosystems. When models are continuously updated on newly collected samples, they can suffer from catastrophic forgetting (CF)~\cite{french1999catastrophic,rahman2022limitations,rahman2025madar}: knowledge of previously observed threats is overwritten by patterns learned from recent data. This issue becomes more severe at the operational scale, where more than 450,000 new malware and potentially unwanted applications (PUA) are observed each day~\cite{av-test}, and large-scale analysis such as VirusTotal processes millions of samples daily~\cite{virustotal}.

Continual learning (CL) offers a framework for malware detection, allowing models to adapt to newly emerging or evolving samples without storing or retraining on the full history of observed  data~\cite{wang2026rethinking,lillo2026activation,jo2026memoryfree,malcl,rahman2025madar}.
However, applying CL to malware detection introduces distinct challenges. Malware data often exhibits complex feature representations, high class imbalance, and a continual stream of malware variants, which makes CL methods for computer vision difficult to adopt directly~\cite{rahman2022limitations}.

\begin{figure}[!t]
\centering

\begin{minipage}{0.5\textwidth}
    \centering
    \includegraphics[width=\linewidth] {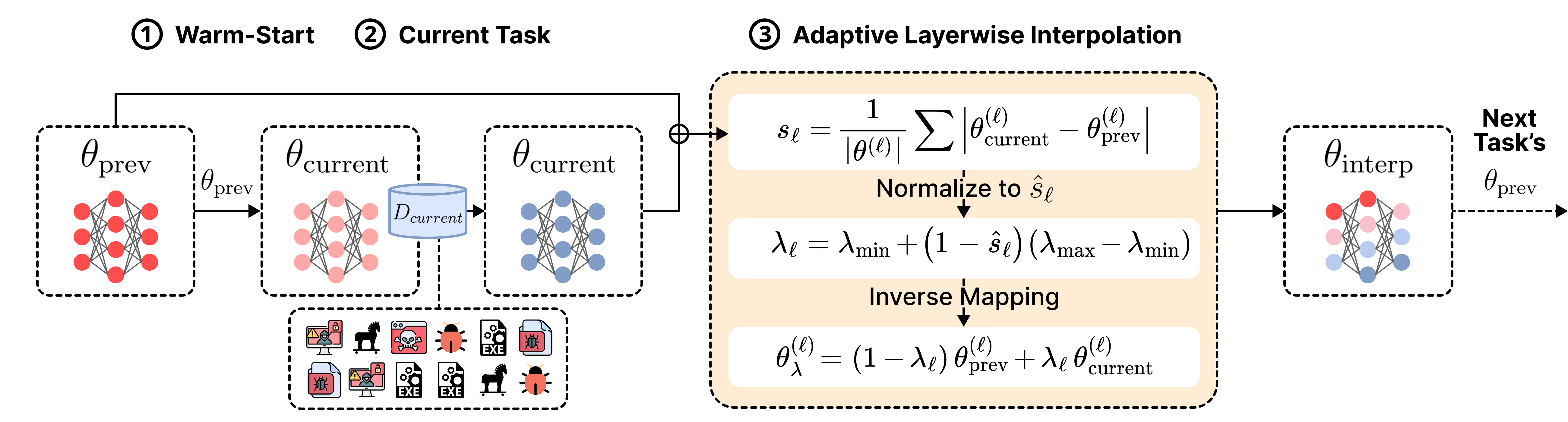}
    \caption{\system learns malware tasks sequentially using warm-starting and adaptive layer-wise interpolation, with weights derived from parameter shifts to enable replay-free stability--plasticity balancing.}
    
    \label{fig:FreeMOCA}
\end{minipage}
\hfill
\begin{minipage}{0.48\textwidth}
    \centering
    \resizebox{\linewidth}{!}{
        \begin{tikzpicture}[
    scale=0.95,
    every node/.style={font=\small},
    optimum/.style={
        circle,
        draw=black,
        fill=white,
        thick,
        minimum size=6pt,
        inner sep=0pt
    },
    interp/.style={
        star,
        star points=5,
        draw=black,
        fill=white,
        thick,
        minimum size=9pt,
        inner sep=0pt
    },
    basin/.style={draw=gray!45, thick},
    basinlight/.style={draw=gray!25, thick},
    lowpath/.style={thick},
    badpath/.style={thick, dashed},
    arrowpath/.style={->, thick, >=Stealth}
]

\begin{scope}[xshift=0cm]
\node[font=\bfseries] at (0,2.25) {(A) Independent training};

\fill[panelablue!8] (-1.45,0) ellipse (1.45 and 0.95);
\fill[panelablue!8] (1.45,0) ellipse (1.45 and 0.95);
\draw[basin, draw=panelablue!75!black] (-1.45,0) ellipse (1.0 and 0.65);
\draw[basinlight, draw=panelablue!45] (-1.45,0) ellipse (1.45 and 0.95);
\draw[basin, draw=panelablue!75!black] (1.45,0) ellipse (1.0 and 0.65);
\draw[basinlight, draw=panelablue!45] (1.45,0) ellipse (1.45 and 0.95);
\draw[panelablue!30, thick] (0,0) ellipse (2.8 and 1.45);

\node[optimum, draw=panelablue!75!black, fill=white] (a1) at (-1.45,0) {};
\node[optimum, draw=panelablue!75!black, fill=white] (a2) at (1.45,0) {};
\node[below] at (-1.45,-0.12) {$\theta_a$};
\node[below] at (1.45,-0.12) {$\theta_b$};

\draw[badpath, draw=badred, very thick] (a1) -- (a2);
\draw[lowpath, draw=panelablue!80!black]
    plot[smooth, tension=0.75]
    coordinates {(-1.45,0) (-0.75,0.72) (0,0.95) (0.75,0.72) (1.45,0)};

\node[align=center] at (0,1.34) {curved low-loss path};
\node[align=center, text=badred!90!black] at (0,-1.13) {direct interpolation\\crosses high loss};
\node[align=center] at (0,-1.78) {\footnotesize no shared trajectory};

\draw[rounded corners, panelablue!60] (-3.10,-2.05) rectangle (3.10,2.45);
\end{scope}

\begin{scope}[xshift=6.6cm]
\node[font=\bfseries] at (0,2.25) {(B) Warm-started CL};

\fill[panelbgreen!10] (0,0.1) ellipse (1.75 and 1.05);
\draw[basin, draw=panelbgreen!75!black] (0,0.1) ellipse (1.15 and 0.75);
\draw[basinlight, draw=panelbgreen!45] (0,0.1) ellipse (1.75 and 1.05);
\draw[panelbgreen!30, thick] (0,0.1) ellipse (2.35 and 1.35);

\node[optimum, draw=panelbgreen!75!black, fill=white] (b1) at (-0.55,0.0) {};
\node[optimum, draw=panelbgreen!75!black, fill=white] (b2) at (0.65,0.25) {};
\node[below left] at (-0.55,0.0) {$\theta_{t-1}$};
\node[above right] at (0.65,0.25) {$\theta_t$};

\draw[arrowpath, draw=panelbgreen!80!black] (b1) -- (b2);

\node[align=center] at (0.08,1.27) {low-loss linear path};
\node[align=center] at (0,-1.23) {warm-start preserves\\locality and alignment};
\node[align=center] at (0,-1.78) {\footnotesize consecutive task optima};

\draw[rounded corners, panelbgreen!60] (-2.95,-2.05) rectangle (2.95,2.45);
\end{scope}

\begin{scope}[xshift=13.2cm]
\node[font=\bfseries] at (0,2.25) {(C) \system consolidation};

\fill[panelcorange!10] (0,0.1) ellipse (1.75 and 1.05);
\draw[basin, draw=panelcorange!80!black] (0,0.1) ellipse (1.15 and 0.75);
\draw[basinlight, draw=panelcorange!45] (0,0.1) ellipse (1.75 and 1.05);
\draw[panelcorange!30, thick] (0,0.1) ellipse (2.35 and 1.35);

\node[optimum, draw=panelcorange!80!black, fill=white] (c1) at (-0.55,0.0) {};
\node[optimum, draw=panelcorange!80!black, fill=white] (c2) at (0.65,0.25) {};
\node[interp, draw=panelcorange!90!black, fill=panelcorange!20] (cl) at (0.10,0.13) {};

\node[below left] at (-0.55,0.0) {$\theta_{t-1}$};
\node[above right] at (0.65,0.25) {$\theta_t$};
\node[below right] at (0.10,0.13) {$\theta_\lambda$};

\draw[lowpath, draw=panelcorange!70!black] (c1) -- (c2);
\draw[arrowpath, draw=panelcorange!85!black] (c1) -- (cl);
\draw[arrowpath, draw=panelcorange!85!black] (c2) -- (cl);

\node[align=center] at (0,-1.15) {replay-free consolidation\\direct interpolation};
\node[align=center] at (0,-1.75) {\footnotesize $\theta_{t-1}\rightarrow\theta_t\rightarrow\theta_{t+1}$};

\draw[rounded corners, panelcorange!65] (-2.95,-2.05) rectangle (2.95,2.45);
\end{scope}

\begin{scope}[yshift=-2.75cm, xshift=5.2cm]
\draw[lowpath, draw=gray!70!black] (-5.8,0) -- (-5.2,0);
\node[anchor=west] at (-5.05,0) {\footnotesize low-loss path / interpolation};

\draw[badpath, draw=badred, very thick] (0.075,0) -- (-0.3,0);
\node[anchor=west] at (-0.0005,0) {\footnotesize high-loss linear path};

\node[optimum, draw=gray!70!black, fill=white] at (3.9,0) {};
\node[anchor=west] at (4.1,0) {\footnotesize task optimum};

\node[interp, draw=gray!70!black, fill=gray!15] at (6.7,0) {};
\node[anchor=west] at (6.9,0) {\footnotesize interpolated model};
\end{scope}

\end{tikzpicture} 
    }
    \caption{LMC in warm-started CL. (A) Independent solutions may face high-loss barriers. (B) Warm-starting keeps consecutive task solutions close, enabling low-loss interpolation. (C) \system uses this property for replay-free consolidation.}
    \label{fig:mode_connectivity}
\end{minipage}

\vspace{-0.5cm}
\end{figure}

Recent malware-specific CL approaches, such as MalCL~\cite{malcl} and MADAR~\cite{rahman2025madar}, address this problem through replay. MalCL uses a Generative Adversarial Network (GAN)-based replay generator, while MADAR introduces distribution-aware replay to preserve knowledge of previously seen threats. 
These methods achieve strong performance, but their reliance on replay introduces non-trivial computational and memory costs. This limits their practicality in deployment settings where resources are constrained, such as endpoint security systems and mobile devices. Even approaches that use a small exemplar buffer require selecting, storing, and repeatedly processing representative samples from past tasks, which can become a significant bottleneck in practice~\cite{cho2025memory}.

Our goal is to support continual adaptation to newly emerging and evolving malware while avoiding the deployment costs of replay-based CL, such as storing past samples, generating synthetic samples, and repeatedly training on them. 
To this end, we propose {\system}, an interpolation-based CL framework for malware detection that requires neither exemplar buffers nor generative replay~\cite{malcl,rahman2025madar}. 


The key idea is to preserve prior knowledge by operating directly in parameter space rather than data space. \system leverages \textit{Linear Mode Connectivity (LMC)}~\cite{garipov2018loss,doan2023continual}, which suggests that different solutions can be connected through low-loss paths. While LMC typically depends on the relationship between solutions, with models from the same initialization being easier to connect linearly than independently trained ones~\cite{wen2023optimizing,zhou2026exploring,mirzadeh2021linear}, CL provides a favorable setting: each task is warm-started from the previous task optimum, inducing implicit alignment between consecutive models. Rather than relying on replay, activation alignment or permutation-based matching, or post-hoc repair~\cite{mirzadeh2020linear,ren2024analyzing,kozal2024continual,tatro2020optimizing,ainsworth2023git}, \system exploits this warm-started structure to interpolate directly between consecutive task optima. This yields a buffer-free mechanism for adapting to new malware samples while retaining prior knowledge. We validate this approach through loss-barrier and variance-collapse diagnostics in Section~\ref{sec:lossbarrier} and \ref{sec:varianceCollapse}.

Direct interpolation is enabled by the warm-started structure, but applying a single mixing coefficient to all parameters can be suboptimal~\cite{zhao2023does,seo2025budgeted}. Different layers change at different rates, with lower layers often serving as relatively stable feature extractors, whereas higher layers adapt more rapidly to new tasks~\cite{zhao2023does,lesort2021continual}. To capture this, \system introduces adaptive layer-wise interpolation, assigning a separate mixing coefficient $\lambda_\ell$ to each layer. This design is motivated by the \textit{Nested Learning} paradigm~\cite{behrouz2025nested}, which models neural networks as multi-frequency systems.


Empirically, \system outperforms 11 baselines by 1–57 percentage points in the Class-Incremental (Class-IL) setting, which is central to malware analysis due to the continual emergence of new malware families. In the Domain-Incremental (Domain-IL) setting, \system remains matching or exceeding five baselines. These results show that exploiting geometric alignment in parameter space provides a scalable, replay-free mechanism for continual malware detection, without storing or revisiting past samples.


Our key contributions are summarized as follows:

\begin{smitemize}
    \item We propose \system, a memory- and compute-efficient CL framework for malware analysis that preserves prior knowledge by interpolating consecutive warm-started task optima, without replay or generative modeling.
    
    \item We analyze LMC in the warm-started CL and show that recursive optimization induces local parameter alignment between consecutive task optima.
    
    \item We introduce {\em Adaptive Layer-wise Interpolation}, which sets layer-specific interpolation weights from parameter shifts to balance stability and plasticity.
    
    \item We evaluate {\system} on large-scale Class-IL malware benchmarks, EMBER-Class and AZ-Class, achieving state-of-the-art average accuracy and outperforming 11 baselines.
    On temporal Domain-IL benchmarks, \system remains competitive, reaching 93.6\% on EMBER-Domain and 97.0\% on AZ-Domain.
    
    \item We provide a comprehensive analysis using forgetting and transfer metrics, along with geometric diagnostics such as loss-barrier and variance-collapse analysis, to explain when and why interpolation remains stable.
\end{smitemize}

\section{Related Work}


\paragraph*{Primary Approaches in CL.}
Replay methods retain prior task information by storing exemplars~\cite{er,icarl}, generating synthetic samples~\cite{bir}, replaying compressed latent activations~\cite{tadros2022sleep}, or distilling soft outputs from earlier models~\cite{lwf}. 
While effective, they often incur additional memory or computational overhead. Regularization-based methods instead preserve prior knowledge by penalizing changes to parameters estimated to be important for earlier tasks. For example, EWC~\cite{ewc} uses the Fisher information matrix, while SI~\cite{si} estimates importance from each parameter’s contribution along the training trajectory. However, such constraints can limit adaptation to new tasks by biasing optimization toward previous solutions. In contrast, \system first allows standard adaptation to the new task and then restores the stability-plasticity balance by interpolating between the previous and current task solutions.


Mean Teacher~\cite{tarvainen2017mean} and test-time adaptation~\cite{wang2022continual} methods improve robustness by averaging weights within a single training trajectory. In contrast, FreeMOCA performs task-level interpolation between consecutive task-specific minima, leveraging warm-started mode connectivity to balance stability and plasticity across tasks. 
Related work on Weight Space Consolidation(WSC) studies replay-rich conditions, combining parameter resetting and weight averaging to reduce GPU time when sufficient exemplar memory is available~\cite{cho2025memory}. \system instead targets a strictly buffer-free setting, using interpolation as the primary consolidation mechanism without replay buffers or generators.

\paragraph*{CL for Malware Domain.}
Research on CL for malware detection is relatively underexplored, and standard CL methods often fail to prevent CF in this domain~\cite{rahman2022limitations}. Recent malware-specific approaches rely mainly on replay: MalCL uses GAN-based generative replay using a feature-matching loss for improving synthetic samples and intermediate classifier features for maintaining class balance~\cite{malcl}. MADAR selects representative replay samples with Isolation Forests in input or feature space or in model features to track evolving malware~\cite{rahman2025madar}. 

\section{FreeMOCA Overview}


\subsection{Problem Setup}
We consider a supervised CL setting in which a classifier $f_\theta$ is trained on a sequence of tasks in $t\in\{1,\ldots,T\}$. Each task $t$ provides a dataset $D_t = \{(x_i, y_i)\}_{i=1}^{N_t}$ sampled from a task-specific distribution. The objective is to minimize the error across tasks, defined as $\arg \min_\theta \frac{1}{T} \sum_{t=1}^T \mathbb{E}_{(x,y) \sim D_t}[\mathcal{L}(f_\theta(x), y)]$. 

Training proceeds sequentially--for each task $t$, the model is initialized from the previous solution $\theta_{t-1}$ and optimized on $D_t$ to obtain $\theta_t$. To consolidate knowledge, \system applies an interpolation operator $\varphi$ between consecutive solutions, \(\theta_t\leftarrow \varphi (\theta_{t-1}, \theta_t)\), where $\varphi$ performs layer-wise interpolation with weights determined by parameter shifts. The resulting model is then used to initialize the next task. 

The theoretical foundations of the LMC assumption in CL, warm-starting, and adaptive layer-wise interpolation are detailed in Sections~\ref{sec:modeconnect}, \ref{sec:warm_start}, and \ref{sec:adaptiveinterpole}. The overall procedure is summarized in Figure~\ref{fig:FreeMOCA} and Algorithm~\ref{alg:freemoca}.
% ...

\begin{algorithm}[H]
\caption{\system with adaptive layer-wise interpolation.}
\label{alg:freemoca}
\begin{algorithmic}[1]
\STATE $\theta_{\mathrm{prev}} \gets \theta_0$
\FOR{$t\gets1$ \textbf{to} $T$}{
    \STATE $\theta_{\mathrm{cur}} \gets \theta_{\mathrm{prev}}$
    \STATE \texttt{//~Init.~from $\theta_{\mathrm{prev}}$ train on $D_t$ to get $\theta_{\mathrm{cur}}$}
    \IF{$t>1$}
        \FOR{layer $\ell$}
            \STATE $s_\ell \gets \frac{1}{|\theta^{(\ell)}|}\|\theta_{\mathrm{cur}}^{(\ell)}-\theta_{\mathrm{prev}}^{(\ell)}\|_1$
        \ENDFOR
        \STATE $\hat{s}_\ell \gets \frac{s_\ell-s_{\min}}{s_{\max}-s_{\min}+\epsilon}$
        \FOR{layer $\ell$}
            \STATE $\lambda_\ell \gets \lambda_{\min}+(1-\hat{s}_\ell)(\lambda_{\max}-\lambda_{\min})$
            \STATE $\theta_t^{(\ell)} \gets (1-\lambda_\ell)\theta_{\mathrm{prev}}^{(\ell)}+\lambda_\ell\theta_{\mathrm{cur}}^{(\ell)}$
        \ENDFOR
    \ELSE
        \STATE $\theta_t \gets \theta_{\mathrm{cur}}$
    \ENDIF
    \STATE $\theta_{\mathrm{prev}} \gets \theta_t$
}
\ENDFOR
\end{algorithmic}
\end{algorithm}

\subsection{Mode Connectivity in CL}
\label{sec:modeconnect}

\system is grounded in LMC, which studies whether two solutions can be joined by a low-loss linear path:
\begin{equation}
\theta_\lambda = (1-\lambda)\theta_a + \lambda \theta_b, \quad \lambda \in [0,1].
\end{equation}
Deep-network minima are often connected by low-loss paths, particularly when trained from shared initializations or related trajectories~\cite{garipov2018loss,draxler2018essentially,frankle2020linear}. In CL, this setting arises naturally: each task is warm-started from the previous solution, making the key question whether consecutive task optima remain sufficiently aligned for interpolation to preserve performance.


\paragraph{Key Observation}
We empirically observe that consecutive warm-started task optima often exhibit low-loss interpolation paths (Section~\ref{sec:lossbarrier}), suggesting that \emph{local linear connectivity} emerges naturally along the CL trajectory.

\paragraph{Proposition 1 (Local Linear Connectivity)}
Let $\theta_{t-1}$ and $\theta_t$ be consecutive task solutions, and let $L_t$ be the loss for task $t$. If (i) task drift is bounded (i.e., $\|\nabla L_t(\theta_{t-1})\|$ is small) and (ii) $L_t$ is twice continuously differentiable with bounded Hessian norm, then interpolation remains within a low-loss neighborhood:
\begin{equation}
L_t(\theta_\lambda)
\le
(1-\lambda)L_t(\theta_{t-1}) + \lambda L_t(\theta_t)
+ \mathcal{O}(\|\theta_t-\theta_{t-1}\|^2).
\end{equation}

This follows from a second-order Taylor expansion around $\theta_{t-1}$:
\begin{equation}
L_t(\theta_{t-1}+\Delta)
\approx
L_t(\theta_{t-1})
+
\nabla L_t(\theta_{t-1})^\top \Delta
+
\frac{1}{2}\Delta^\top H_t(\theta_{t-1})\Delta,
\label{eq:taylor}
\end{equation}
where $\Delta=\theta_t-\theta_{t-1}$ and $H_t$ is the Hessian~\cite{bottou2018optimization,martens2010deep}. Under bounded curvature and small task drift, the path between consecutive solutions stays locally with low loss. This interpretation is consistent with basin reuse in transfer learning~\cite{neyshabur2020being} and prior work on weight interpolation in CL~\cite{kozal2024continual}. It is also supported by evidence that different layers evolve non-uniformly across tasks~\cite{zhao2023does}.

Taken together, these observations suggest that warm-started optimization induces \emph{local geometric continuity} across tasks. We therefore view CL as a \emph{progressive chain of locally connected solutions}, where each task yields an optimum that remains sufficiently aligned with its predecessor for interpolation to be effective. Formally, we express consolidation as
\begin{equation}
\theta_t \leftarrow \phi(\theta_{t-1}, \theta_t),
\end{equation}
where $\phi$ denotes the interpolation operator that merges the previous solution with the current update.

Importantly, this is a \emph{local} claim: \system leverages the empirically supported property that warm-started optimization preserves sufficient alignment between adjacent tasks for interpolation to remain effective. We validate this hypothesis using loss-barrier and variance-collapse analyses (Sections~\ref{sec:lossbarrier} and~\ref{sec:varianceCollapse}).

\subsection{Warm-Started Connectivity}
\label{sec:warm_start}

\system does not assume that arbitrary task optima are globally connected by low-loss linear paths. Instead, our claim is explicitly \emph{local}: interpolation is reliable primarily between \emph{consecutive warm-started solutions}. Warm-starting induces two key properties: \emph{locality} and \emph{alignment}. 

\paragraphB{Locality}
Because each task is initialized from the previous optimum, gradient-based updates produce a solution $\theta_t$ that remains within a nearby region of the loss landscape under moderate task drift. Under standard smoothness assumptions, updates satisfy
\begin{equation}
\theta_t \approx \theta_{t-1} - \eta \nabla L_t(\theta_{t-1}),
\end{equation}
implying that the displacement $\Delta = \theta_t - \theta_{t-1}$ remains small when gradients are bounded. This is consistent with observations from transfer learning, where fine-tuning preserves basin membership when tasks share structure~\cite{neyshabur2020being}, and from federated learning, where averaging models within the same basin yields stable solutions~\cite{mcmahan2017communication}.

\paragraphB{Alignment}
Warm-starting also preserves parameter correspondence. Since $\theta_t$ is obtained by continuing optimization from $\theta_{t-1}$, parameters maintain consistent semantics across tasks: neuron or channel $i$ at task $t$ is a continuation of its counterpart at task $t-1$. This distinguishes \system from interpolation methods that combine independently trained models, where misaligned representations can degrade performance.

These properties directly support the local connectivity assumption in Section~\ref{sec:modeconnect}: locality ensures that consecutive solutions lie within the same low-loss region, while alignment ensures that interpolation operates over corresponding parameters. This also explains why explicit neuron permutation is unnecessary. While permutation symmetry complicates interpolation across independently trained networks, warm-started CL preserves parameter correspondence by construction. This is consistent with regularization-based CL methods that constrain parameters by name and index without explicit matching~\cite{kirkpatrick2017overcoming,si}.

We emphasize that this is a scoped claim: explicit alignment may still be useful when interpolating models from different seeds or architectures, but for consecutive warm-started solutions, direct interpolation is the natural and effective operation. The overview of LMC with warm-start is illustrated in Figure~\ref{fig:mode_connectivity}.

\subsection{Adaptive Layer-wise Interpolation}
\label{sec:adaptiveinterpole}

While Sections~\ref{sec:modeconnect}--\ref{sec:warm_start} establish that consecutive task optima are locally connected, this connectivity is not uniform across layers. A uniform interpolation coefficient is therefore often suboptimal. Different layers encode features at varying abstraction levels and exhibit distinct transferability properties~\cite{yosinski2014transferable,raghu2019transfusion}, adapting at different rates during CL.
Applying a single $\lambda$ ignores this heterogeneity, potentially over-regularizing adaptive layers or under-constraining stable ones.

To address this, \system employs a layer-wise interpolation strategy where each layer $\ell$ is assigned a distinct coefficient $\lambda_\ell$:
\begin{equation}
\theta_{\lambda}^{(\ell)} = (1 - \lambda_\ell)\theta_{\text{prev}}^{(\ell)} + \lambda_\ell \theta_{\text{current}}^{(\ell)}.
\end{equation}
Here, $\lambda_\ell \in [\lambda_{\min}, \lambda_{\max}]$ determines the influence of the current model. To capture this heterogeneity, we quantify layer \emph{plasticity} via the magnitude of parameter updates:
\[
s_\ell = \frac{\|\theta_{\text{current}}^{(\ell)} - \theta_{\text{prev}}^{(\ell)}\|_1}{|\theta^{(\ell)}|}.
\]
Mapping $s_\ell$ to the range $[0, 1]$ via min-max normalization yields $\hat{s}_\ell$, from which we derive:
\begin{equation}
\lambda_\ell = \lambda_{\min} + (1 - \hat{s}_\ell)(\lambda_{\max} - \lambda_{\min}).
\end{equation}

This mechanism acts as an adaptive regularizer: layers undergoing significant updates (high $\hat{s}_\ell$) are assigned smaller $\lambda_\ell$ to prioritize stability, while stable layers (low $\hat{s}_\ell$) are assigned larger $\lambda_\ell$ to favor task adaptation. This balances stability and plasticity by adapting to the non-uniform update dynamics across layers observed in CL~\cite{zhao2023does,qiao2024learn}.

\section{Experimental Details}
\label{sec:experiment_details}

\paragraphB{Datasets} We use four large-scale benchmarks adapted from~\cite{rahman2022limitations, rahman2025madar}, covering both Windows (EMBER~\cite{ember}) and Android (AndroZoo~\cite{AndroZoo}) environments. In the main paper, we focus on the Class-IL setting using EMBER-Class (337,035 samples; $>$400 per family) and AZ-Class (285,582 samples; $>$200 per family), each comprising 100 distinct malware families.
For EMBER, we utilize 2,381-dimensional hashed vectors capturing PE/header metadata and function calls. For AZ, features are extracted via DREBIN~\cite{arp2014drebin} across eight semantic categories (e.g., API calls and network addresses), yielding 2,439-dimensional feature vectors. 

We also conduct Domain-IL experiments on EMBER-Domain and AZ-Domain; the corresponding setup and results are provided in Appendix~\ref{app:domainil}.

\paragraphB{Model and Metrics}
We use a three-layer Vanilla CNN as the default backbone in our main experiments, allowing us to evaluate the proposed interpolation mechanism independently of architectural complexity. The model is optimized via SGD with learning rate $10^{-3}$, momentum 0.9, and weight decay $10^{-7}$. To further evaluate the architectural robustness of \system, we include three alternative backbones in our ablation study (Section~\ref{app:ablation}): a Vision Transformer (ViT), a deeper CNN, and a Convolutional Kolmogorov--Arnold Network (CKAN)~\cite{liu2024kan}. See Appendix~\ref{app:interpolation} for details on interpolation methods.


To comprehensively assess \system, we report {\em Average Accuracy}, {\em Forgetting ($F$)}, and {\em Remembering (REM)}~\cite{diaz2018don}. Knowledge transfer is evaluated using {\em Forward Transfer (FWT)} and {\em Backward Transfer (BWT/BWT$^{+}$)}. Details of all metrics are provided in Appendix~\ref{app:metric}.

\paragraphB{Baseline and Comparison}
We benchmark against two boundary conditions: \textit{None} (fine-tuning only; lower bound) and \textit{Joint} (offline training on cumulative data; upper bound). For comparative analysis in Class-IL, we evaluate 11 established CL methods: EWC~\cite{ewc}, SI~\cite{si}, LwF~\cite{lwf}, GR, ER~\cite{er}, iCaRL~\cite{icarl}, BI-R~\cite{van2020brain}, MalCL~\cite{malcl}, TAMiL~\cite{tamil}, CLeWI~\cite{kozal2024continual}, and WSC~\cite{cho2026forgetforgettingcontinuallearning}. For Domain-IL, we restrict evaluation to EWC, SI, GR, MalCL, and LwF.

\section{Evaluation}
\label{sec:evaluation}

\begin{table*}[!t]
\centering
\scriptsize
\caption{{\bf \system in Class-IL}: Average accuracy on EMBER-Class and AZ-Class, compared with prior work (\textbf{L}: Linear, \textbf{S}: Spline, \textbf{P}: Polynomial).}
\label{tab:mean_results_msr}

\setlength{\tabcolsep}{1.25pt}
\resizebox{\textwidth}{!}{
\begin{tabular}{c|cc|ccccccccccc|ccc|ccc}
\toprule
\multirow{3}{*}{\textbf{Dataset}} 
& \multicolumn{2}{c|}{\textbf{Baseline}} 
& \multicolumn{11}{c|}{\textbf{Prior Work}} 
& \multicolumn{6}{c}{\textbf{FreeMOCA (Ours)}} \\
\cline{2-20}

& \multirow{2}{*}{\textbf{None}} 
& \multirow{2}{*}{\textbf{Joint}} 
& \multirow{2}{*}{\textbf{EWC}} 
& \multirow{2}{*}{\textbf{SI}}  
& \multirow{2}{*}{\textbf{GR}} 
& \multirow{2}{*}{\textbf{MalCL}} 
& \multirow{2}{*}{\textbf{LwF}} 
& \multirow{2}{*}{\textbf{TAMiL}} 
& \multirow{2}{*}{\textbf{ER}} 
& \multirow{2}{*}{\textbf{iCaRL}} 
& \multirow{2}{*}{\textbf{BI-R}} 
& \multirow{2}{*}{\textbf{CLeWI}} 
& \multirow{2}{*}{\textbf{WSC}} 
& \multicolumn{3}{c|}{\textbf{Fixed $\lambda = 0.6$}}  
& \multicolumn{3}{c}{\textbf{Adap. Layerwise $\lambda$}} \\ 
\cline{15-20}

& & & & & &
& & & & & & &
& L & S & P & L & S & P \\ 

\midrule

EMBER 
& 22.9
& 88.5
& 25.6  
& 8.7   
& 26.6  
& 54.5  
& 10.7  
& 33.2  
& 26.9  
& 55.6  
& 26.8  
& 60.1  
& 64.1  
& \textbf{65.2} 
& \textbf{65.2} 
& \textbf{65.2} 
& \textbf{66.3} 
& \textbf{63.5} 
& \textbf{63.8} \\

\midrule

AZ 
& 26.5
& 86.1
& 17.1  
& 9.3   
& 22.9  
& 59.0  
& 10.6  
& 54.8  
& 56.9  
& 54.7  
& 22.5  
& 62.2  
& 55.7  
& \textbf{63.7} 
& \textbf{63.7} 
& \textbf{63.7} 
& \textbf{66.1} 
& \textbf{63.9} 
& \textbf{64.2} \\ 

\bottomrule
\end{tabular}
}
\end{table*}

\begin{table*}[t]
\centering
\begin{minipage}[t]{0.605\textwidth}
    \centering
    \caption{CL transfer and forgetting metrics for EMBER-Class and AZ-Class (all values reported as fractions).}

\tiny
\centering
\setlength{\tabcolsep}{1.5pt}
\begin{tabular}{llccccc}
\toprule
\textbf{Dataset} & \textbf{Method} & \textbf{FWT ($\uparrow$)} & \textbf{BWT ($\uparrow$)} & \textbf{BWT$^{+}$ ($\uparrow$)} & \textbf{F ($\downarrow$)} & \textbf{REM ($\uparrow$)} \\
\midrule
\multirow{10}{*}{\rotatebox[origin=c]{90}{EMBER}}   
                    & FreeMOCA & \textbf{0.0002} & \textbf{-0.0060} & 0.0000 & \textbf{0.0060} & \textbf{0.9940} \\
                    & EWC      & 0.0000 & -0.4059 & 0.0000 & 0.4482 & 0.5941 \\
                    & SI       & 0.0000 & -0.8534 & 0.0000 & 0.9447 & 0.1465 \\
                    & GR       & 0.0000 & -0.6040 & 0.0000 & 0.6469 & 0.3960 \\
                    & MalCL    & 0.0000 & -0.1477 & 0.0000 & 0.1507 & 0.8523 \\
                    & LwF      & 0.0000 & -0.1691 & 0.0000 & 0.0958 & 0.8309 \\
                    & TAMiL    & 0.0000 & -0.2569 & 0.0000 & 0.5334 & 0.7431 \\
                    & iCaRL    & 0.0000 & -0.0986 & 0.0000 & 0.1470 & 0.9013 \\
                    & CLeWI    & 0.0000 & -0.1444 & 0.0000 & 0.0625 & 0.8555 \\
                    & WSC      & 0.0000 & -0.2144 & 0.0000 & 0.2052 & 0.7856 \\
\midrule
\multirow{10}{*}{\rotatebox[origin=c]{90}{AZ}}   
                    & FreeMOCA & 0.0000 & \textbf{0.1480} & \textbf{0.1480} & \textbf{0.0000} & \textbf{1.0000} \\
                    & EWC      & 0.0000 & -0.6549 & 0.0000 & 0.5680 & 0.3450 \\
                    & SI       & 0.0000 & -0.8463 & 0.0000 & 0.8498 & 0.1536 \\
                    & GR       & 0.0000 & -0.5264 & 0.0000 & 0.6138 & 0.4735 \\
                    & MalCL    & 0.0000 & -0.1753 & 0.0000 & 0.2563 & 0.8246 \\
                    & LwF      & 0.0000 & -0.1625 & 0.0000 & 0.0900 & 0.8374 \\
                    & TAMiL    & 0.0000 & -0.2948 & 0.0000 & 0.3793 & 0.7051 \\
                    & iCaRL    & 0.0000 & -0.1270 & 0.0000 & 0.1806 & 0.8729 \\
                    & CLeWI    & 0.0000 &  0.0144 & 0.0144 & 0.0589 & \textbf{1.0000} \\
                    & WSC      & 0.0000 & -0.3397 & 0.0000 & 0.3770 & 0.6603 \\
\bottomrule
\end{tabular}

    \label{tab:forgetting_metrics}
\end{minipage}
\hfill
\begin{minipage}[t]{0.335\textwidth}
    \centering
    \caption{Final$\rightarrow$final loss-barrier heights $B_{t\rightarrow t+1}$ (Linear path) for AZ-Class and EMBER-Class.}
    \footnotesize
\centering
\resizebox{\linewidth}{!}{
\begin{tabular}
{
  >{\centering\arraybackslash}p{1.35cm} |
  >{\centering\arraybackslash}p{1.35cm} |
  >{\centering\arraybackslash}p{1.35cm}
}
\toprule
\textbf{Transition} $(t\rightarrow t+1)$ 
& \textbf{AZ-Class} $B_{t\rightarrow t+1}$ 
& \textbf{EMBER-Class} $B_{t\rightarrow t+1}$ \\
\midrule
$0\rightarrow 1$  & 0.2579 & 0.1462 \\
$1\rightarrow 2$  & 0.1292 & 0.0741 \\
$2\rightarrow 3$  & 0.0147 & 0.0328 \\
$3\rightarrow 4$  & 0.0530 & 0.0013 \\
$4\rightarrow 5$  & 0.0109 & 0.0153 \\
$5\rightarrow 6$  & 0.0295 & 0.0185 \\
$6\rightarrow 7$  & 0.0260 & 0.0033 \\
$7\rightarrow 8$  & 0.0204 & 0.0008 \\
$8\rightarrow 9$  & 0.0248 & 0.0026 \\
$9\rightarrow 10$ & 0.0188 & 0.0027 \\
\midrule
Mean $\pm$ Std & 0.0585 $\pm$ 0.0742 & 0.0298 $\pm$ 0.0444 \\
Median         & 0.0254             & 0.0093 \\
Max            & 0.2579             & 0.1462 \\
\bottomrule
\end{tabular}
}

    \label{tab:barriers_linear_two_datasets}
\end{minipage}
\end{table*}

\subsection{Class-Incremental (Class-IL) Setting}
\label{sec:class_il}

Consistent with prior Class-IL settings~\cite{rahman2022limitations,malcl,rahman2025madar}, we partition the dataset into 11 sequential tasks: an initial task (Task 0) with 50 randomly selected malware families, followed by 10 tasks (Task 1--10) each introducing five new families, progressively expanding the label space. In Class-IL, where the classifier head expands as new classes are introduced, \system interpolates only dimension-matched parameters; newly added FC weights are copied directly from the current-task model rather than interpolated. For FreeMOCA, we evaluate both adaptive layer-wise and fixed-$\lambda$ settings across linear, spline, and polynomial interpolation. Empirically, $\lambda=0.6$ yields the most robust performance across datasets. Sensitivity analyses are provided in the ablation study (Section~\ref{app:ablation}).

\paragraphB{Overall Performance with Global Accuracy}
We adopt a compact CNN backbone to evaluate FreeMOCA under a controlled setting. As shown in Table~\ref{tab:mean_results_msr}, the \textit{Joint} baseline achieves strong upper-bound performance (88.5\% on EMBER-Class and 86.1\% on AZ-Class), while the \textit{None} baseline degrades sharply to 22.9\% and 26.5\%, respectively, highlighting severe CF. FreeMOCA substantially recovers this gap, reaching 65.2\% on EMBER-Class and up to 66.1\% on AZ-Class. Compared to replay-based methods such as iCaRL, ER, and TAMiL, it consistently improves performance by approximately 9-11 percentage points. While the margin over the strongest baselines (e.g., WSC on EMBER-Class) is modest, FreeMOCA achieves competitive or superior performance without storing past data. This demonstrates that parameter-space interpolation can effectively replace replay in memory- and privacy-constrained settings.

\paragraphB{Analysis of Transfer and Forgetting}
As shown in Table~\ref{tab:forgetting_metrics}, \system consistently outperforms all baselines across forgetting and retention metrics. On EMBER-Class, it achieves the lowest forgetting ($F=0.006$) and the highest remembering (REM = 0.994), while maintaining near-zero backward transfer (BWT = -0.006), indicating almost complete preservation of prior knowledge. On AZ-Class, \system exhibits a markedly stronger effect: it achieves zero forgetting ($F=0.000$) and perfect retention (REM = 1.000), while yielding substantial positive backward transfer (BWT = 0.148). Although CLeWI shows marginal positive BWT (0.014), \system is the only method with consistently strong positive backward transfer, indicating that learning new malware families can meaningfully improve performance on previously learned tasks. These results demonstrate that \system not only mitigates CF but can induce beneficial backward transfer, all without storing past data, relying solely on geometric interpolation in parameter space.

\begin{figure*}[t]
\centering

\begin{minipage}[t]{0.505\textwidth}
\centering
\begin{subfigure}[t]{0.46\textwidth}
\centering
\begin{tikzpicture}
\begin{axis}[
    ybar,
    axis lines=left,
    width=4.cm,
    height=3.2cm,
    ymin=60,
    ymax=66,
    symbolic x coords={B1,B2,B1+B2,FC,FreeMOCA},
    xtick=data,
    x tick label style={font=\tiny, rotate=60, anchor=east},
    yticklabel style={font=\scriptsize},
    ylabel={Accuracy (\%)},
    ylabel style={font=\scriptsize, yshift=-0.45cm},
    enlarge x limits=0.12,
    nodes near coords,
    every node near coord/.append style={font=\tiny, rotate=90, anchor=west},
]
\addplot coordinates {
    (B1,63.3)
    (B2,61.0)
    (B1+B2,62.1)
    (FC,63.16)
    (FreeMOCA,65.2)
};
\end{axis}
\end{tikzpicture}
\caption{EMBER-Class}
\label{fig:block-exp-bar-ember}
\end{subfigure}
\hfill
\begin{subfigure}[t]{0.46\textwidth}
\centering
\begin{tikzpicture}
\begin{axis}[
    ybar,
    axis lines=left,
    width=4.cm,
    height=3.2cm,
    ymin=60,
    ymax=66,
    symbolic x coords={B1,B2,B1+B2,FC,FreeMOCA},
    xtick=data,
    x tick label style={font=\tiny, rotate=60, anchor=east},
    yticklabel style={font=\scriptsize},
    enlarge x limits=0.12,
    nodes near coords,
    every node near coord/.append style={font=\tiny, rotate=90, anchor=west},
]
\addplot coordinates {
    (B1,63.3)
    (B2,62.7)
    (B1+B2,63.2)
    (FC,62.8)
    (FreeMOCA,63.7)
};
\end{axis}
\end{tikzpicture}
\caption{AZ-Class}
\label{fig:block-exp-bar-az}
\end{subfigure}

\caption{Selective block interpolation analysis with fixed $\lambda=0.5$. Interpolating only Block~1 yields the highest accuracy on both datasets, suggesting early layers, which capture more transferable low-level representations, are better suited for interpolation than deeper, more task-specific layers.} 
\label{fig:block-exp-bar}
\end{minipage}
\hfill
\begin{minipage}[t]{0.475\textwidth}
\centering
\subcaptionbox{EMBER-Class\label{fig:loss-barrier-ember}}[0.49\linewidth]{%
    \includegraphics[width=\linewidth]{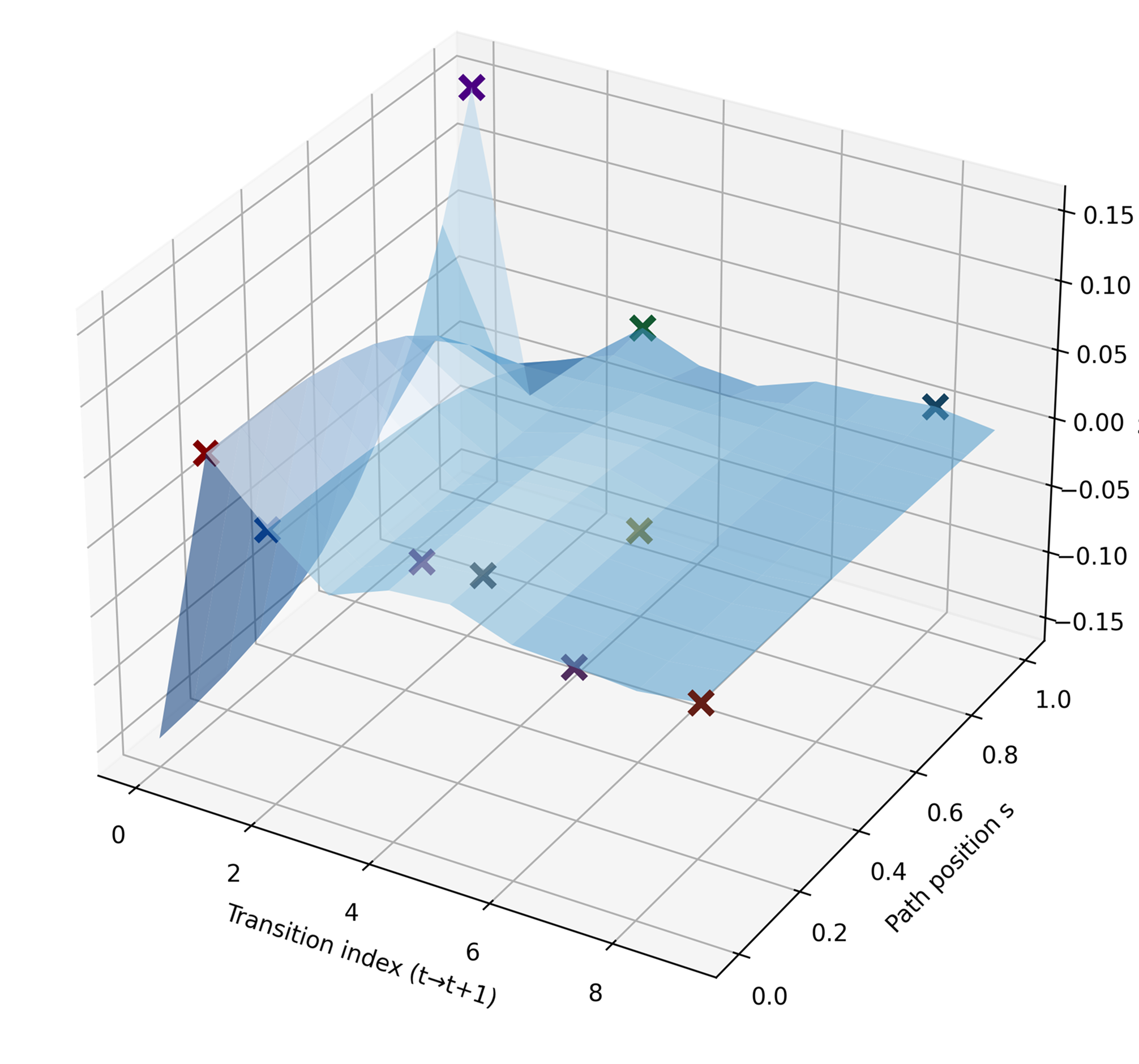}
}
\hfill
\subcaptionbox{AZ-Class\label{fig:loss-barrier-az}}[0.49\linewidth]{%
    \includegraphics[width=\linewidth]{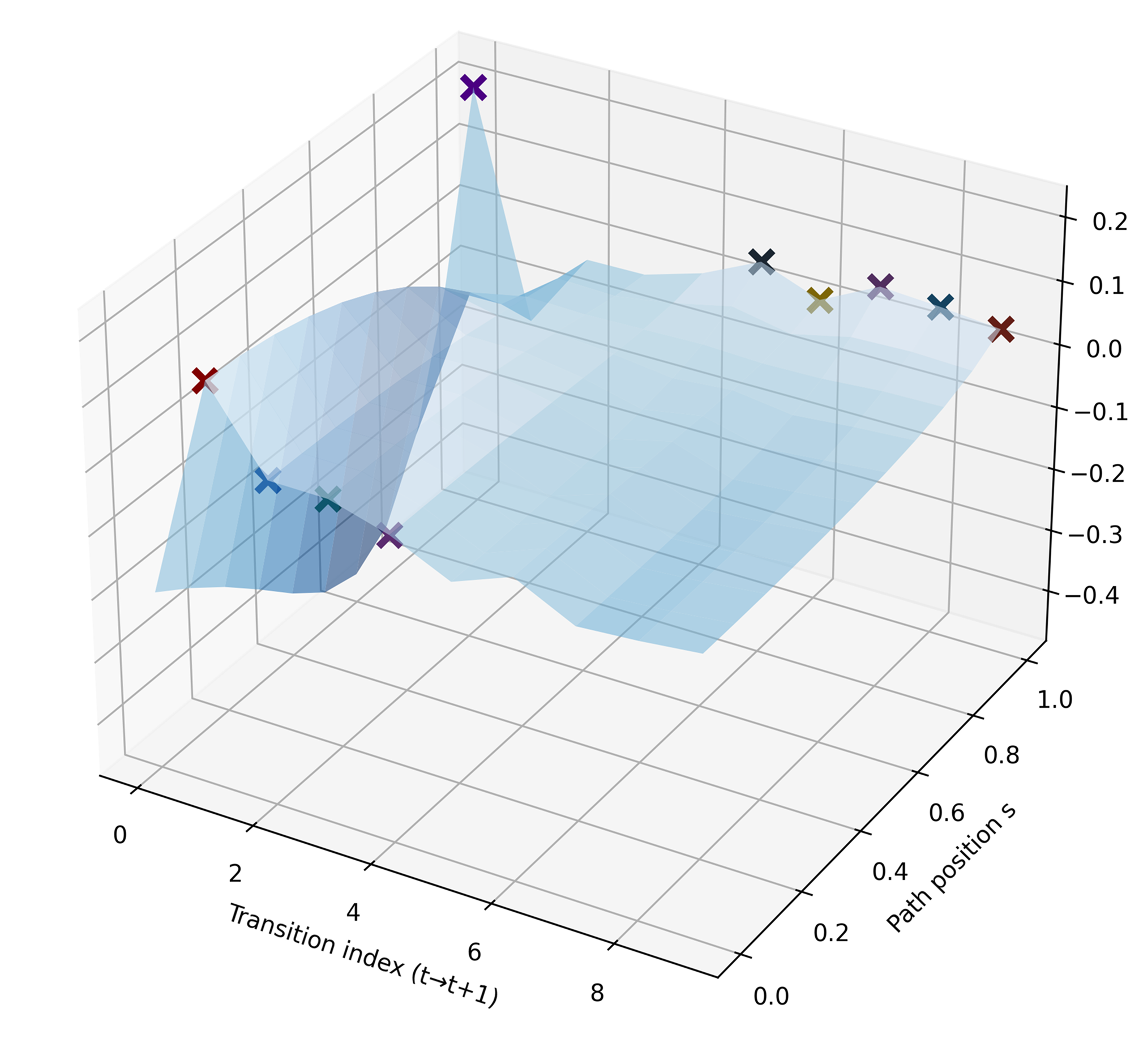}
}
\caption{Chain of connected valleys in the loss-barrier surface across CL transitions (interpolated model at $t\!-\!1 \rightarrow t$). After the initial transition, the barriers remain flat, indicating that \system stays within a locally connected low-loss region.}
\label{fig:loss-barrier-combined}
\end{minipage}

\end{figure*}

\subsection{Loss-Barrier Diagnostics for Mode Connectivity}
\label{sec:lossbarrier}

To verify whether \system maintains optimization within a connected low-loss region, we analyze the loss landscape between consecutive solutions via \emph{loss barriers}~\cite{garipov2018loss,frankle2020linear}, defined as the maximum deviation along a linear interpolation path from the interpolation of endpoint losses.

\paragraphB{Loss Barrier Definition}
Given two consecutive solutions $\theta_a$ and $\theta_b$, we define the linear path
\begin{equation}
\theta(s)=(1-s)\theta_a+s\theta_b,\qquad s\in[0,1].
\end{equation}
We quantify the barrier height as
\begin{equation}
B(\theta_a,\theta_b)
=
\max_{s\in[0,1]}
\left[
L(\theta(s)) -
\big((1-s)L(\theta_a)+sL(\theta_b)\big)
\right].
\end{equation}
In practice, we approximate the maximum using a uniform grid of $s$ values. A barrier close to zero indicates approximate LMC between the two solutions.

\paragraphB{Landscape Analysis}
Table~\ref{tab:barriers_linear_two_datasets} and Figure~\ref{fig:loss-barrier-combined} show the evolution of barrier heights across task transitions. We observe a consistent two-phase pattern across both datasets. First, the initial transition exhibits the largest barrier, with $B_{0\rightarrow1}=0.2579$ on AZ-Class and $B_{0\rightarrow1}=0.1462$ on EMBER-Class, reflecting substantial feature reorganization during early task expansion. Second, barrier heights decrease sharply and stabilize at low values. On EMBER-Class, transitions after Task 3 are nearly flat (e.g., $B_{3\rightarrow4}=0.0013$, $B_{7\rightarrow8}=0.0008$), indicating near-linear connectivity. AZ-Class shows a similar trend with slightly higher variability, including occasional moderate spikes (e.g., $B_{3\rightarrow4}=0.0530$), but maintains low barriers overall (mean $0.0585$, median $0.0254$).

These results indicate that, after an initial reorganization phase, \system guides optimization into a locally connected low-loss region, where subsequent task updates occur along near-linear paths.

\begin{table*}[!t]
\tiny
\caption{{\bf Class-IL}: Comparison of fixed and Adaptive Layerwise $\lambda$ on the EMBER-Class and AZ-Class datasets  using different Interpolation (\textbf{Int.}) Methods (\textbf{L}: Linear, \textbf{S}: Spline, \textbf{P}: Polynomial).}
\label{tab:ember_vs_AZ}
\centering
\setlength{\tabcolsep}{2.5pt}
\renewcommand{\arraystretch}{1.2}
\begin{tabular}{l|l|ccccccccc|c}
\toprule
\multirow{2}{*}{\textbf{\rotatebox[origin=c]{90}{Model}}} 
& \multirow{2}{*}{\textbf{\rotatebox[origin=c]{90}{Int.}}} 
& \multicolumn{9}{c|}{$\lambda$} & 
\textbf{Adaptive}
\\ \cline{3-11}
& 
& \textbf{0.1} & \textbf{0.2} & \textbf{0.3} & \textbf{0.4} & \textbf{0.5} & \textbf{0.6} & \textbf{0.7} & \textbf{0.8} & \textbf{0.9}
&  \textbf{Layerwise $\lambda$}\\
\midrule

\multirow{3}{*}{\rotatebox[origin=c]{90}{EMBER}} 
& L 
& 51.1{\tiny $\pm$0.01} & 51.1{\tiny $\pm$0.01} & 51.3{\tiny $\pm$0.01} & 55.2{\tiny $\pm$1.90} & 64.9{\tiny $\pm$0.55} & \textbf{65.2{\tiny $\pm$0.01}} & 64.7{\tiny $\pm$0.12} & 63.5{\tiny $\pm$0.20} & 62.6{\tiny $\pm$0.13}
& 66.3{\tiny $\pm$0.00} \\

& S 
& 51.1{\tiny $\pm$0.02} & 51.1{\tiny $\pm$0.01} & 51.3{\tiny $\pm$0.02} & 57.2{\tiny $\pm$2.00} & 64.5{\tiny $\pm$1.20} & \textbf{65.2{\tiny $\pm$0.00}} & 64.7{\tiny $\pm$0.15} & 63.8{\tiny $\pm$0.20} & 62.8{\tiny $\pm$0.20}
& 63.5{\tiny $\pm$0.00} \\

& P 
& 51.1{\tiny $\pm$0.01} & 51.1{\tiny $\pm$0.07} & 51.2{\tiny $\pm$0.03} & 54.4{\tiny $\pm$1.64} & 64.5{\tiny $\pm$1.20} & \textbf{65.2{\tiny $\pm$0.04}} & 64.7{\tiny $\pm$0.08} & 63.5{\tiny $\pm$0.31} & 62.5{\tiny $\pm$0.16}
& 63.8{\tiny $\pm$0.00} \\

\midrule

\multirow{3}{*}{\rotatebox[origin=c]{90}{AZ}} 
& L     
& 62.5{\tiny $\pm$0.06} & 62.4{\tiny $\pm$0.03} & 62.9{\tiny $\pm$0.04} & 63.4{\tiny $\pm$0.03} & \textbf{63.7{\tiny $\pm$0.01}} & \textbf{63.7{\tiny $\pm$0.03}} & 63.5{\tiny $\pm$0.03} & 63.3{\tiny $\pm$0.03} & 63.1{\tiny $\pm$0.03}
& \textbf{66.1{\tiny $\pm$0.00}} \\

& S     
& 62.4{\tiny $\pm$0.04} & 62.4{\tiny $\pm$0.01} & 62.9{\tiny $\pm$0.01} & 63.4{\tiny $\pm$0.07} & \textbf{63.7{\tiny $\pm$0.07}} & \textbf{63.7{\tiny $\pm$0.11}} & 63.6{\tiny $\pm$0.04} & 63.2{\tiny $\pm$0.07} & 63.0{\tiny $\pm$0.11}
& \textbf{63.9{\tiny $\pm$0.00}} \\

& P 
& 62.4{\tiny $\pm$0.05} & 62.4{\tiny $\pm$0.04} & 62.9{\tiny $\pm$0.02} & 63.4{\tiny $\pm$0.04} & \textbf{63.7{\tiny $\pm$0.04}} & \textbf{63.7{\tiny $\pm$0.11}} & 63.6{\tiny $\pm$0.05} & 63.2{\tiny $\pm$0.06} & 63.0{\tiny $\pm$0.08}
& \textbf{64.2{\tiny $\pm$0.00}} \\

\bottomrule
\end{tabular}
\end{table*}

\subsection{Feature Variance and Batch Normalization Stability}
\label{sec:varianceCollapse}

\begin{wraptable}[12]{r}{0.56\columnwidth}
\vspace{-10pt}
\centering
\scriptsize
\setlength{\tabcolsep}{2pt}
\caption{Variance collapse diagnostics at \texttt{block2}.}
\label{tab:variance_diagnostics_ember_az}
\resizebox{0.56\columnwidth}{!}{
\tiny
\label{tab:variance_diagnostics_ember_az}
\setlength{\tabcolsep}{3pt}
\begin{tabular}{c|cccc|cccc}
\toprule
& \multicolumn{4}{c|}{\textbf{EMBER-Class}} 
& \multicolumn{4}{c}{\textbf{AZ-Class}} \\
\cmidrule(lr){2-5} \cmidrule(lr){6-9}
\textbf{Task} 
& Var$_{\text{prev}}$ & Var$_{\text{curr}}$ & Ratio & Diagnosis
& Var$_{\text{prev}}$ & Var$_{\text{curr}}$ & Ratio & Diagnosis \\
\midrule

1  & 0.1450 & 0.3241 & 1.0376 & No Collapse 
   & 0.3107 & 0.7059 & 1.1509 & No Collapse \\

2  & \cellcolor{gray!15}0.0595 & \cellcolor{gray!15}0.1981 & \cellcolor{gray!15}0.7239 & \cellcolor{gray!15}Collapse 
   & \cellcolor{gray!15}0.1350 & \cellcolor{gray!15}0.5838 & \cellcolor{gray!15}0.5913 & \cellcolor{gray!15}Collapse \\

3  & 0.1045 & 0.2008 & 0.8109 & No Collapse 
   & \cellcolor{gray!15}0.1847 & \cellcolor{gray!15}0.4786 & \cellcolor{gray!15}0.6786 & \cellcolor{gray!15}Collapse \\

4  & 0.1750 & 0.2654 & 0.8594 & No Collapse 
   & \cellcolor{gray!15}0.2339 & \cellcolor{gray!15}0.5726 & \cellcolor{gray!15}0.7188 & \cellcolor{gray!15}Collapse \\

5  & 0.1280 & 0.1922 & 0.9292 & No Collapse 
   & \cellcolor{gray!15}0.3553 & \cellcolor{gray!15}0.5664 & \cellcolor{gray!15}0.7709 & \cellcolor{gray!15}Collapse \\

6  & 0.1946 & 0.2173 & 0.9842 & No Collapse 
   & 0.3724 & 0.6145 & 0.8857 & No Collapse \\

7  & 0.1726 & 0.1836 & 0.9708 & No Collapse 
   & 0.3401 & 0.4588 & 0.8917 & No Collapse \\

8  & 0.1768 & 0.1704 & 0.9856 & No Collapse 
   & 0.3745 & 0.5105 & 0.9151 & No Collapse \\

9  & 0.1994 & 0.1638 & 1.0188 & No Collapse 
   & 0.4357 & 0.5216 & 0.9591 & No Collapse \\

10 & 0.2078 & 0.1921 & 0.9560 & No Collapse 
   & 0.4866 & 0.5250 & 0.9481 & No Collapse \\

\midrule
\multicolumn{1}{l|}{\textbf{Mean Ratio}} 
& \multicolumn{4}{c|}{0.928}
& \multicolumn{4}{c}{0.852} \\

\multicolumn{1}{l|}{\textbf{\#-of Collapses}} 
& \multicolumn{4}{c|}{1 / 10}
& \multicolumn{4}{c}{4 / 10} \\

\bottomrule
\end{tabular}
}
\vspace{-8pt}
\end{wraptable}
Naive weight interpolation can induce \emph{variance collapse}, reducing activation variance and invalidating BatchNorm (BN) statistics~\cite{ioffe2015batch}.
Standard remedies such as BN re-estimation or neuron permutation require access to past data, making them incompatible with FreeMOCA's buffer-free setting. Instead, we hypothesize that warm-starting preserves sufficient alignment between consecutive solutions to maintain stable feature statistics under interpolation~\cite{yosinski2014transferable,raghu2019transfusion}.

To quantify this effect, we define the \emph{Variance Collapse Ratio} $R$, computed on a probe batch at the output of Block~2:
\begin{equation}
R=\frac{\operatorname{Var}(\Phi(\theta_{\text{interp}}))}{(1-\lambda)\operatorname{Var}(\Phi(\theta_{\text{prev}}))+\lambda\operatorname{Var}(\Phi(\theta_{\text{curr}}))}.
\end{equation}
Values of $R\approx 1$ indicate preserved feature geometry, while lower values indicate variance collapse.

As shown in Table~\ref{tab:variance_diagnostics_ember_az}, both datasets exhibit an initial transient followed by stabilization. On EMBER-Class, variance collapse occurs only at Task~2 ($R\approx0.72$), after which the ratio increases (Tasks~3--4, $\approx0.81$--$0.86$) and remains close to unity from Task~5 onward ($\approx0.93$--$1.02$), indicating stable BN behavior. Overall, EMBER exhibits only 1 collapse out of 10 transitions, with a high mean ratio of 0.928. AZ-Class shows a stronger early collapse (Tasks~2--5, $\approx0.59$--$0.77$), but similarly stabilizes from Task~6 ($\approx0.89$--$0.96$). Despite 4 collapses in early transitions, the ratio consistently approaches unity in later stages (mean 0.852), indicating recovery of stable feature statistics. These results indicate that warm-started interpolation mitigates variance collapse and preserves BN stability without requiring re-estimation or permutation.

\subsection{Memory and Computational Efficiency}

\begin{table}[t]
\scriptsize
\centering
\caption{Memory and training-time comparison of MalCL, WSC, and \system.}
\label{tab:memory_cost}
\setlength{\tabcolsep}{2.5pt}
\begin{tabular}{l|c|c|c}
\toprule
\multirow{2}{*}{\textbf{Evaluation}}    & \multicolumn{3}{c}{\textbf{Methods}} \\
\cmidrule(lr){2-4}
                               & \textbf{MalCL} & \textbf{WSC} & \textbf{\system} \\
\midrule
Training Time Avg. per Epoch   & 3804.38\,$s$ & 679.30\,$s$ & 49.00\,$s$ \textbf{(92.79--98.71\% \textcolor{green}{$\downarrow$})} \\
Memory Avg. per Task           & 284.29\,MB & 16.44\,MB & 15.98\,MB \textbf{(2.80--94.38\% \textcolor{green}{$\downarrow$})} \\
Memory Increase ($t=0\rightarrow10$) & 86.3\% \textcolor{red}{$\uparrow$} (223.36$\rightarrow$416.20\,MB) & 45.4\% \textcolor{red}{$\uparrow$} (10.90$\rightarrow$19.99\,MB) & \textbf{0\% $\uparrow$}  (15.98$\rightarrow$15.98\,MB) \\
\bottomrule
\end{tabular}
\end{table}

We compared the practical resource cost of \system with that of the best-performing replay-based and interpolation method, MalCL and WSC. As shown in Table~\ref{tab:memory_cost}, \system achieves a 2.80--94.38\% reduction in memory usage by retaining only a single classifier checkpoint from the previous task. In contrast, MalCL demonstrated substantially higher memory overhead, including a generator/classifier checkpoint (207.43\,MB and 15.90\,MB), synthetic replay memory (4.65\,MB), and replay buffers that grow with the task sequence (20.98\,MB at $t=1$ to 188.83\,MB at $t=10$).  WSC demonstrated an increasing burden of memory cost (9.08\,MB at $t=1$ to 19.16\,MB at $t=10$) due to the usage of the memory buffer. \system exhibits substantially lower training time than MalCL and WSC, achieving a 92.79--98.7\% reduction in per-epoch training time. Overall, \system attains higher detection accuracy and lower forgetting than competitors, while requiring significantly less memory and computation.

\section{Ablation Study}
\label{app:ablation}

\paragraphB{Selective Block Interpolation}

\begin{wraptable}[9]{r}{0.53\columnwidth}
\vspace{-25pt}
\centering
\scriptsize
\setlength{\tabcolsep}{2pt}
\caption{Robustness analysis of FreeMOCA accuracy on EMBER-Class across different seeds, optimizers, learning-rate schedules, and backbones.}
\resizebox{0.54\columnwidth}{!}{
\tiny
\centering
\vspace{-0.1cm}
\setlength{\tabcolsep}{3.5pt}
\label{tab:robust_results}
\begin{tabular}{lclclclc}
\toprule
\multicolumn{8}{c}{\textbf{\system}} \\
\midrule
\multicolumn{2}{c}{\textbf{Seed}} & \multicolumn{2}{c}{\textbf{Optimizers}} & \multicolumn{2}{c}{\textbf{LR Schedules}} & \multicolumn{2}{c}{\textbf{Backbone}} \\
\cmidrule(lr){1-2}
\cmidrule(lr){3-4}
\cmidrule(lr){5-6}
\cmidrule(lr){7-8}
Seed 1,2  & 64-65 & SGD      & 69 & StepLR            & 65 & ViT           & 60 \\
Seed 3  & 66 & Adam     & 62 & MultiStepLR       & 65 & Deeper CNN    & 57 \\
Seed 4,5  & 69 & AdamW    & 64 & CosineAnnealingLR & 68 & Conv. KAN     & 52 \\
\bottomrule
\end{tabular}


}
\end{wraptable}

To analyze how network depth affects mode connectivity, we interpolate selected blocks while fixing $\lambda = 0.5$. As shown in Figure~\ref{fig:block-exp-bar}, the full \system model achieves the highest accuracy on both datasets, whereas interpolating only Block~1 gives the best result among the selective-block settings, yielding 63.28\% on EMBER-Class and 63.33\% on AZ-Class. This is consistent with the hierarchical structure of CNNs: earlier layers tend to capture more generic and transferable features, making them better candidates for interpolation. In contrast, interpolating deeper layers such as Block~2 reduces performance, reaching only 60.96\% on EMBER-Class, likely because deeper layers encode more task-specific representations. Moreover, interpolating both Block~1 and Block~2 does not improve over Block~1 alone, suggesting that selective interpolation of early layers is more effective than interpolating deeper or multiple blocks.

\paragraphB{Sensitivity Analysis of $\lambda$}
\label{app:lambda_results}
We examine the plasticity--stability trade-off by varying the interpolation weight $\lambda$ across linear, spline, and polynomial interpolation methods.
As shown in Table ~\ref{tab:ember_vs_AZ}, both datasets under the class-incremental setting exhibit a consistent trend: accuracy improves as $\lambda$ increases, peaks at $\lambda =0.6$, and then gradually declines. This phenomenon holds across all three interpolation methods. At $\lambda=0.6$, EMBER-Class achieves the highest accuracy of 65.2\%, while the AZ dataset reaches 63.7\%. 
Smaller values of $\lambda$ help preserve prior knowledge but limit adaptation to new classes. In contrast, larger values of $\lambda$ bias the model toward the updated model, improving plasticity. Overall, the range $0.4\leq\lambda\leq0.6$ provides the balance between retaining knowledge from prior tasks and adapting to new ones. 
Importantly, our adaptive-layerwise-$\lambda$ method achieves comparable or better performance than fixed $\lambda$ settings. This demonstrates that dynamically learning $\lambda$ is an effective alternative to manual tuning.

\begin{wrapfigure}{r}{0.53\columnwidth}
    \centering
    \vspace{-14pt}
    \adjustbox{width=0.54\textwidth}{
        \begin{tikzpicture}
\begin{axis}[
    width=0.9\textwidth,
    height=0.42\textwidth,
    xlabel={Incremental Classes},
    ylabel={Accuracy (\%)},
    xmin=0, xmax=10,
    ymin=45, ymax=95,
    xtick={0,1,2,3,4,5,6,7,8,9,10},
    ymajorgrids=true,
    grid style={dashed,gray!30},
    tick label style={font=\bf,yshift=2,xshift=-2},
    label style={font=\bf},
    clip=false,
    legend style={
        draw=black,
        fill=white,
        rounded corners=2pt,
        at={(0.97,0.97)},   
        anchor=north east
    },
    legend cell align={left},
]

\addplot[
    red,
    thick,
    mark=*,
] coordinates {
    (0,93.0316)
    (1,80.5318)
    (2,78.2251)
    (3,71.3706)
    (4,65.3390)
    (5,63.1227)
    (6,61.3297)
    (7,60.0204)
    (8,54.4165)
    (9,51.6915)
    (10,50.4718)
};
\addlegendentry{EMBER, avg. = 66.3\%}

\addplot[
    blue,
    thick,
    mark=square*,
] coordinates {
    (0,89.12)
    (1,83.99)
    (2,79.06)
    (3,73.45)
    (4,69.08)
    (5,65.11)
    (6,56.00)
    (7,55.10)
    (8,53.05)
    (9,49.70)
    (10,49.36)
};
\addlegendentry{AZ-Class, avg. = 66.1\%}

\end{axis}
\end{tikzpicture}
    }
    \caption{Task-wise classification accuracy across incremental classes under the Class-IL setting for EMBER-Class and AZ-Class.}
    \label{fig:ember-az-task-wise-accuracy}
    \vspace{-12pt}
\end{wrapfigure}
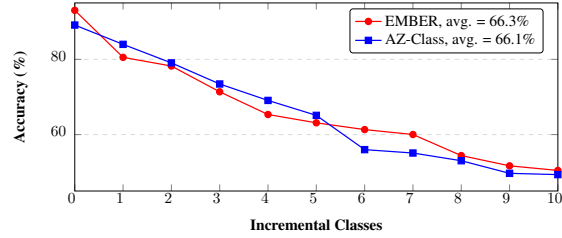


Figure~\ref{fig:ember-az-task-wise-accuracy} shows the task-wise accuracy of \system on EMBER-Class and AZ-Class. Both curves decline as the number of learned classes grows, highlighting the increasing difficulty of maintaining prior knowledge over longer sequences.

\paragraphB{FreeMOCA Hyperparameters} 
We explored various random seeds, optimizers, learning-rate schedules, and backbone architectures, as summarized in Table~\ref{tab:robust_results}. Across five seeds and three optimizers, \system consistently achieved stable performance, indicating that it is relatively insensitive to initialization and optimizer choices. Using SGD as the optimizer, we further compare three learning-rate schedulers and observe only marginal differences in performance.
We also replace the default backbone architecture with three alternatives: a ViT with seven Transformer encoder layers and a CLS token, a deeper CNN with ten convolutional layers, and a Convolutional KAN composed of three ConvKAN layers followed by a KAN linear layer. Across all architectures, \system substantially improves over the \textit{None} baseline, increasing accuracy from 18.23\% to 60.43\% (ViT), 16.02\% to 56.66\% (Deeper CNN), and 13.42\% to 52.01\% (Convolutional KAN). These results suggest that the effectiveness of \system remains robust across various architectural choices.

\section{Conclusion}
In this work, we propose a memory-free CL framework for malware classification that mitigates CF  without storing historical data. By leveraging mode connectivity, \system constructs low-loss interpolation paths between task-specific models, enabling effective knowledge retention across evolving tasks. Experimental results demonstrate that \system consistently outperforms existing CL methods on Windows and Android malware benchmarks in the CL and DL settings.

\bibliographystyle{plainnat}
\bibliography{main-reference}


\appendix

\section{Discussion and Limitations}
\label{Discussioon-Limitation}
\system is designed for a specific CL regime: sequential tasks with observable boundaries and sufficient representational continuity between adjacent updates. Its central assumption is local rather than global connectivity. We do not claim that arbitrary task optima are universally connected by low-loss linear paths; instead, the method relies on the empirically supported observation that warm-started optimization often preserves local alignment across neighboring tasks.

Consequently, performance depends on the degree of feature reuse between consecutive tasks. When task drift becomes large or the underlying representation changes substantially, interpolation quality may deteriorate as adjacent solutions no longer remain within the same locally connected region. Similarly, because FreeMOCA recursively consolidates models through repeated interpolation, approximation errors may accumulate over long task horizons.

The method also assumes explicit task transitions and therefore does not directly address fully task-free or streaming CL settings. While our experiments include stress tests beyond the primary malware benchmarks, the strongest empirical evidence currently lies in structured evolving malware classification, where task progression naturally preserves partial feature continuity. Broader generalization beyond such regimes should therefore be interpreted cautiously.

Despite these limitations, the results suggest that local geometric connectivity provides a viable alternative to replay-based consolidation under memory and privacy constraints. More broadly, our findings indicate that CL may benefit from viewing sequential optimization trajectories as connected regions of the loss landscape rather than isolated task-specific optima.

Future work includes extending FreeMOCA to task-free online learning, studying interpolation under stronger distribution shifts, and developing theoretically grounded adaptive interpolation dynamics for large-scale and multi-modal CL systems.

\section{Interpolation Methods}
\label{app:interpolation}
To operationalize the geometric connectivity discussed earlier, we seek to construct a continuous trajectory function, $\mathcal{F}(x)$, that passes through a given set of discrete data points, $(x_i, y_i)$. Formally, this requires the interpolant to satisfy the constraint $\mathcal{F}(x_i) = y_i$ for every task solution in the sequence. The mathematical formulation of this path dictates not only the smoothness of the transition but also whether the trajectory remains within the low-loss manifold.
We consider three primary approaches: Linear, Polynomial, and Spline Interpolation.

\paragraphB{Linear Interpolation (LI)} LI is the simplest interpolation method that approximates the function between consecutive data points using straight-line segments. It estimates an unknown value by connecting its two closest known data points with a straight line. 


\paragraphB{Polynomial Interpolation (PI)}
PI captures the curvature of the loss landscape by fitting a global polynomial of degree $n$ across all $n$+1 data points. 
While PI yields a globally smoother interpolant than LI, it is limited by its global nature, where changes to any single data point affect the entire interpolating curve and the propensity for Runge's Phenomenon, which causes severe non-physical oscillations between data points for a large number of points or poorly distributed data.

\paragraphB{Spline Interpolation (SLI)}
SLI mitigates the instability of global polynomials by leveraging \emph{piecewise low-degree polynomials}, typically cubic, to construct the path. Instead of a single volatile curve, SLI stitches together local segments while enforcing continuity of low-order derivatives at the connection points (knots), slope ($\mathcal{F}'$), and curvature ($\mathcal{F}''$) at every knot. This locality is crucial: it prevents the catastrophic oscillations seen in PI, ensuring the trajectory remains physically plausible and stable within the loss region even as more tasks are added.






\section{Definitions of Metrics}
\label{app:metric}

In this section, we define the transferability and forgetting evaluation metrics. Following the established protocol, we construct a per-task performance matrix $R \in \mathbb{R}^{N \times N}$, where each entry $R_{i,j}$ denotes the test accuracy on task $j$ after completing training on task $i$.

\paragraph{Forward Transfer (FWT).}
Forward transfer measures how much learning earlier tasks improves performance on future tasks before those tasks are trained (i.e., zero-shot generalization). It is defined as the average of the upper-triangular entries of $R$ (excluding the diagonal):
\begin{equation}
\mathrm{FWT}=\frac{2}{N(N-1)}\sum_{i<j} R_{i,j}.
\end{equation}

\begin{table*}[!t]
\scriptsize
\caption{{\bf Domain-IL}: Comparison of fixed and Adaptive Layerwise $\lambda$ with CNN classifier on the EMBER-Domain and AZ-Domain datasets using different Interpolation (\textbf{Int.}) Methods (\textbf{L}: Linear, \textbf{S}: Spline, \textbf{P}: Polynomial).}
\label{tab:ember_vs_AZ_domain}
\centering
\setlength{\tabcolsep}{2.5pt}
\renewcommand{\arraystretch}{1.2}
\begin{tabular}{l|l|ccccccccc|c}
\toprule
\multirow{2}{*}{\textbf{\rotatebox[origin=c]{90}{Model}}} 
& \multirow{2}{*}{\textbf{\rotatebox[origin=c]{90}{Int.}}} 
& \multicolumn{9}{c|}{$\lambda$} & 
\textbf{Adaptive}
\\ \cline{3-11}
& 
& \textbf{0.1} & \textbf{0.2} & \textbf{0.3} & \textbf{0.4} & \textbf{0.5} & \textbf{0.6} & \textbf{0.7} & \textbf{0.8} & \textbf{0.9}
&  \textbf{Layerwise $\lambda$}\\
\midrule

\multirow{3}{*}{\rotatebox[origin=c]{90}{EMBER}} 
& L 
& 92.2{\scriptsize $\pm$0.40} & 92.7{\scriptsize $\pm$0.20} & 92.7{\scriptsize $\pm$ 0.40} & 92.4{\scriptsize $\pm$ 0.10} & 92.7{\scriptsize $\pm$0.50} & 92.6{\scriptsize $\pm$ 0.20} & 92.9{\scriptsize $\pm$ 0.40} & 93.0{\scriptsize $\pm$0.60} & 92.4{\scriptsize $\pm$0.70} & 93.0{\scriptsize $\pm$0.10} \\

& S 
& 93.5{\scriptsize $\pm$0.13} & 93.5{\scriptsize $\pm$0.14} & 93.4{\scriptsize $\pm$0.01} & 93.5{\scriptsize $\pm$0.12} & 93.3{\scriptsize $\pm$0.17} & 93.4{\scriptsize $\pm$0.24} & 93.4{\scriptsize $\pm$0.06} & 93.6{\scriptsize $\pm$0.17} & 93.0{\scriptsize $\pm$0.10} & 92.2{\scriptsize $\pm$0.50} \\

& P 
& 93.5{\scriptsize $\pm$0.05} & 93.5{\scriptsize $\pm$0.06} & 93.4{\scriptsize $\pm$0.13} & 93.4{\scriptsize $\pm$0.14} & 93.6{\scriptsize $\pm$0.14} & 93.6{\scriptsize $\pm$0.09} & 93.5{\scriptsize $\pm$0.22} & 93.5{\scriptsize $\pm$0.11} & 93.4{\scriptsize $\pm$0.06}
& 93.3{\scriptsize $\pm$0.31} \\

\midrule

\multirow{3}{*}{\rotatebox[origin=c]{90}{AZ}} 
& L     
& 97.0{\scriptsize $\pm$0.03} & 97.0{\scriptsize $\pm$0.03} & 97.0{\scriptsize $\pm$0.01} & 97.0{\scriptsize $\pm$0.03} & 97.0{\scriptsize $\pm$0.02} & 97.0{\scriptsize $\pm$0.05} & 97.0{\scriptsize $\pm$0.04} & 97.0{\scriptsize $\pm$0.05} & 97.0{\scriptsize $\pm$0.03}
& 97.0{\scriptsize $\pm$0.03} \\

& S     
& 97.0{\scriptsize $\pm$0.04} & 97.0{\scriptsize $\pm$0.02} & 97.0{\scriptsize $\pm$0.01} & 97.0{\scriptsize $\pm$0.02} & 97.0{\scriptsize $\pm$0.00} & 97.0{\scriptsize $\pm$0.00} & 97.0{\scriptsize $\pm$0.04} & 97.0{\scriptsize $\pm$0.01} & 97.0{\scriptsize $\pm$0.01}
& 97.0{\scriptsize $\pm$0.00} \\

& P 
& 97.0{\scriptsize $\pm$0.02} & 97.0{\scriptsize $\pm$0.05} & 97.0{\scriptsize $\pm$0.01} & 97.0{\scriptsize $\pm$0.03} & 97.0{\scriptsize $\pm$0.08} &  97.0{\scriptsize $\pm$0.03} & 97.0{\scriptsize $\pm$0.05} & 97.0{\scriptsize $\pm$0.05} & 97.0{\scriptsize $\pm$0.02}
& 97.0{\scriptsize $\pm$0.03} \\ 

\bottomrule
\end{tabular}
\end{table*}

\paragraph{Backward Transfer (BWT).}
Backward transfer captures the effect of learning new tasks on the performance of past tasks. It compares accuracy on a previous task $j$ after learning a later task $i$ to the accuracy achieved immediately after learning task $j$ itself (the diagonal element $R_{j,j}$):
\begin{equation}
\mathrm{BWT}=\frac{2}{N(N-1)}\sum_{i=2}^{N}\sum_{j=1}^{i-1}\left(R_{i,j}-R_{j,j}\right).
\end{equation}
A positive BWT indicates that later learning improves earlier tasks (positive backward transfer), while a negative BWT indicates degradation on past tasks, i.e., forgetting.

\paragraph{Positive Backward Transfer (BWT$^{+}$).}
To isolate improvements on past tasks, we report the non-negative component of BWT:
\begin{equation}
\mathrm{BWT}^{+}=\max(\mathrm{BWT},0).
\end{equation}
This value reflects only beneficial backward transfer, ignoring forgetting.

\paragraph{Forgetting ($F$) and Remembering (REM).}
Following the same decomposition, forgetting is defined as the magnitude of the negative part of BWT:
\begin{equation}
F=\left|\min(\mathrm{BWT},0)\right|=-\min(\mathrm{BWT},0),
\end{equation}
and remembering is reported as a score in $[0,1]$ (higher is better):
\begin{equation}
\mathrm{REM}=1-\left|\min(\mathrm{BWT},0)\right|=1-F.
\end{equation}
Thus, larger $F$ indicates more forgetting, while larger REM indicates better retention of earlier tasks\cite{diaz2018don}.

\section{Domain-Incremental (Domain-IL) Setting}
\label{app:domainil}

\begin{table*}[!t]
\centering
\scriptsize
\caption{{\bf \system in Domain-IL}: Average accuracy on EMBER-Domain and AZ-Domain, compared with prior work (\textbf{L}: Linear, \textbf{S}: Spline, \textbf{P}: Polynomial).}
\label{tab:mean_results_domain}
\setlength{\tabcolsep}{2.5pt}
\begin{tabular}{c|cc|ccccc|ccc|ccc}
\toprule
\multirow{3}{*}{\textbf{Dataset}} & \multicolumn{2}{c|}{\textbf{Baseline}} & \multicolumn{5}{c|}{\textbf{Prior Work}} & \multicolumn{6}{c}{\textbf{FreeMOCA (Ours)}} \\ \cline{2-14}
 & \multirow{2}{*}{\textbf{None}} &
 \multirow{2}{*}{\textbf{Joint}} & 
 \multirow{2}{*}{\textbf{EWC}} & 
 \multirow{2}{*}{\textbf{SI}} & 
 \multirow{2}{*}{\textbf{GR}} & 
 \multirow{2}{*}{\textbf{MalCL}} & 
 \multirow{2}{*}{\textbf{LwF}} & \multicolumn{3}{c|}{\textbf{Fixed $\lambda = 0.6$}}  & \multicolumn{3}{c}{\textbf{Adap. Layerwise $\lambda$}} \\ \cline{9-14}
 & & & & & & &
 & L & S & P & L & S & P \\

\midrule

EMBER & 
92.1 & 
96.5 & 
90.9 & 
91.5 & 
91.8 & 
93.2 & 
93.0 & 
92.6 & 
\textbf{93.4} & 
\textbf{93.6} & 
93.0 & 92.2 & \textbf{93.3}\\

\midrule

AZ & 
96.0 & 
97.8 & 
95.4 & 
94.4 & 
95.5 & 
\textbf{97.0} & 
95.2 & 
\textbf{97.0} & 
\textbf{97.0} & 
\textbf{97.0} & 
\textbf{97.0} & 
\textbf{97.0} & 
\textbf{97.0}\\ 
\bottomrule
\end{tabular}
\end{table*}






\begin{table}[!t]
\scriptsize
\centering
\caption{CL transfer and forgetting metrics for EMBER-Domain and AZ-Domain (all values reported as fractions).}
\label{tab:forgetting_metrics_domain}
\setlength{\tabcolsep}{3pt}
\begin{tabular}{llccccc}
\toprule
\textbf{Dataset} & \textbf{Method} & \textbf{FWT ($\uparrow$)} & \textbf{BWT ($\uparrow$)} & \textbf{BWT$^{+}$ ($\uparrow$)} & \textbf{F ($\downarrow$)} & \textbf{REM ($\uparrow$)} \\
\midrule
\multirow{6}{*}{\centering\rotatebox[origin=c]{90}{EMBER}}
                    & FreeMOCA & \textbf{0.9091} & -0.0335 & 0.0000 & 0.0335 & 0.9664 \\
                    & EWC      & 0.8838 & -0.0594 & 0.0000 & 0.0472 & 0.9406 \\
                    & SI       & 0.8924 & \textbf{-0.0106} & 0.0000 & \textbf{0.0190} &\textbf{ 0.9894} \\
                    & GR       & 0.8857 & -0.0386 & 0.0000 & 0.0486 & 0.9614 \\
                    & MalCL    & 0.9005 & -0.0401 & 0.0000 & 0.0416 & 0.9598 \\
                    & LwF      & 0.8881 & -0.0302 & 0.0000 & 0.0357 & 0.9698 \\
\midrule
\multirow{6}{*}{\centering\rotatebox[origin=c]{90}{AZ}}
                    & FreeMOCA & 0.9160 & -0.0322 & 0.0000 & 0.0322 & 0.9677 \\
                    & EWC      & 0.9165 & \textbf{-0.0103} & 0.0000 & \textbf{0.0128} & \textbf{0.9897} \\
                    & SI       & \textbf{0.9198} & -0.0261 & 0.0000 & 0.0332 & 0.9739 \\
                    & GR       & 0.9182 & -0.0197 & 0.0000 & 0.0229 & 0.9803 \\
                    & MalCL    & 0.9133 & -0.0129 & 0.0000 & 0.0240 & 0.9871 \\
                    & LwF      & 0.9158 & -0.0208 & 0.0000 & 0.0269 & 0.9792 \\
\bottomrule
\end{tabular}
\end{table}

\begin{table}[!t]
\scriptsize
\centering
\caption{Interpolated model at t-1$\rightarrow$interpolated model at t loss-barrier heights $B_{t\rightarrow t+1}$ (Max Spike) for AZ-Domain and EMBER-Domain.}
\label{tab:barriers_maxspike_two_domains}
\begin{tabular}{
>{\centering\arraybackslash}p{1.85cm} |
>{\centering\arraybackslash}p{1.85cm} |
>{\centering\arraybackslash}p{1.85cm}
}
\toprule
Transition $(t\rightarrow t+1)$ & AZ-Domain $B_{t\rightarrow t+1}$ & EMBER-Domain $B_{t\rightarrow t+1}$ \\
\midrule
$0\rightarrow 1$  & 0.2393 & 0.1350 \\
$1\rightarrow 2$  & 0.1137 & 0.0812 \\
$2\rightarrow 3$  & 0.0651 & 0.0452 \\
$3\rightarrow 4$  & 0.0826 & 0.0423 \\
$4\rightarrow 5$  & 0.0705 & 0.0758 \\
$5\rightarrow 6$  & 0.0826 & 0.0201 \\
$6\rightarrow 7$  & 0.0743 & 0.0400 \\
$7\rightarrow 8$  & --      & 0.0546 \\
$8\rightarrow 9$  & --      & 0.1603 \\
$9\rightarrow 10$ & --      & 0.0724 \\
$10\rightarrow 11$& --      & 0.0352 \\
\midrule
Mean $\pm$ Std & 0.1040 $\pm$ 0.057 & 0.0693 $\pm$ 0.04 \\
Median & 0.08263 & 0.05462 \\
Max & 0.2393 & 0.1603 \\
\bottomrule
\end{tabular}
\end{table}

\begin{table}[t]
\centering
\scriptsize
\caption{Variance collapse diagnostics across incremental tasks at the \texttt{block2} layer.}
\label{tab:variance_diagnostics_ember_az_domain}
\setlength{\tabcolsep}{3pt}

\begin{tabular}{c|cccc|cccc}
\toprule
& \multicolumn{4}{c|}{\textbf{EMBER-Domain}} 
& \multicolumn{4}{c}{\textbf{AZ-Domain}} \\
\cmidrule(lr){2-5} \cmidrule(lr){6-9}
\textbf{Task} 
& Var$_{\text{prev}}$ & Var$_{\text{curr}}$ & Ratio & Diagnosis
& Var$_{\text{prev}}$ & Var$_{\text{curr}}$ & Ratio & Diagnosis \\
\midrule

1  & 0.2267 & 0.2525 & 1.0486 & No Collapse 
   & \cellcolor{gray!15}0.1119 & \cellcolor{gray!15}0.2178 & \cellcolor{gray!15}1.3157 & \cellcolor{gray!15}Collapse \\

2  & 0.2330 & 0.2419 & 1.0209 & No Collapse 
   & 0.2139 & 0.2375 & 1.0345 & No Collapse \\

3  & 0.2475 & 0.2631 & 1.0307 & No Collapse 
   & 0.2232 & 0.2466 & 1.0369 & No Collapse \\

4  & 0.2375 & 0.2534 & 1.0257 & No Collapse 
   & 0.2245 & 0.2635 & 1.0634 & No Collapse \\

5  & 0.2387 & 0.2380 & 1.0067 & No Collapse 
   & 0.2227 & 0.2465 & 1.0405 & No Collapse \\

6  & 0.2453 & 0.2522 & 1.0114 & No Collapse 
   & 0.2293 & 0.2439 & 1.0277 & No Collapse \\

7  & 0.2481 & 0.2494 & 1.0093 & No Collapse 
   & 0.2202 & 0.2372 & 1.0290 & No Collapse \\

8  & 0.2396 & 0.2405 & 1.0050 & No Collapse 
   & -- & -- & -- & -- \\

9  & 0.2239 & 0.2493 & 1.0405 & No Collapse 
   & -- & -- & -- & -- \\

10 & 0.2453 & 0.2435 & 1.0054 & No Collapse 
   & -- & -- & -- & -- \\

11 & 0.2293 & 0.2316 & 1.0072 & No Collapse 
   & -- & -- & -- & -- \\

\midrule
\multicolumn{1}{l|}{\textbf{Mean Ratio}} 
& \multicolumn{4}{c|}{1.019}
& \multicolumn{4}{c}{1.078} \\

\multicolumn{1}{l|}{\textbf{\# Collapses}} 
& \multicolumn{4}{c|}{0 / 11}
& \multicolumn{4}{c}{1 / 7} \\

\bottomrule
\end{tabular}
\end{table}

We extend our analysis to the Domain-IL setting to evaluate robustness against temporal concept drift. Unlike Class-IL, Domain-IL preserves a simpler binary decision boundary and already yields strong fine-tuning performance, so this setting provides less headroom for improvement. Therefore, we view Domain-IL as a secondary robustness assessment rather than the primary regime in which the benefits of FreeMOCA are expected to be the most pronounced. We utilize temporal splits to simulate concept drift: EMBER-Domain is partitioned into 12 monthly tasks, while AZ-Domain is split into 9 yearly tasks from 2008 to 2016.
For this setting, we employ the fixed configuration ($\lambda=0.6$) and restrict our comparative baseline to methods compatible with domain shifts: EWC, SI, GR, MalCL, and LwF.




\paragraphB{Results}
In the Domain-IL setting, \system improves upon the \textit{None} baseline by 1.5 and 1.0 percentage points on EMBER-Domain and AZ-Domain, respectively. We note that the performance gains of \system are more limited due to the nature of Domain-IL: unlike Class-IL, decision boundaries remain relatively simpler, resulting in a higher baseline performance with less room for improvement. 
Nevertheless, \system consistently matched or outperformed all baseline models, demonstrating its effectiveness across diverse datasets and settings.




\paragraphB{Transfer and Forgetting Metrics Results}
Compared to the Class-IL setting, all methods achieve substantially improved performance in terms of both transferability and forgetting, as shown in Table~\ref{tab:forgetting_metrics_domain}, which indicates a simpler decision boundary in Domain-IL. 

Table ~\ref{tab:forgetting_metrics_domain} reports consistently high forward transfer across both datasets, with \system achieving the best FWT on EMBER-Domain ($0.9091$) and comparable transfer on AZ-Domain ($0.9160$). Across all methods, BWT is slightly negative (thus $\mathrm{BWT}^+=0$), indicating mild forgetting rather than positive backward transfer. Retention differences are small but consistent: SI yields the lowest forgetting on EMBER ($F=0.0190$, $\mathrm{REM}=0.9894$), while EWC provides the strongest retention on AZ ($F=0.0128$, $\mathrm{REM}=0.9897$). Overall, \system remains strong, exhibiting a modest stability--plasticity trade-off (EMBER: $F=0.0335$, $\mathrm{REM}=0.9664$; AZ: $F=0.0322$, $\mathrm{REM}=0.9677$).

\paragraphB{Loss-Barrier Diagnostics for Mode Connectivity} Table~\ref{tab:barriers_maxspike_two_domains} reports the interpolated model at task $t-1$ to the interpolated model at task $t$ loss-barrier heights $B_{t\rightarrow t+1}$ between consecutive tasks. Barriers are overall low in both settings (AZ: mean $0.1040\pm0.057$, median $0.0826$, max $0.2393$; EMBER: mean $0.0693\pm0.040$, median $0.0546$, max $0.1603$), indicating that successive solutions remain largely connected by low-loss paths. AZ exhibits a higher average barrier than EMBER, suggesting slightly stronger domain shifts across its task sequence, with the largest spikes occurring at $0\rightarrow1$ (AZ) and $8\rightarrow9$ (EMBER).

\paragraphB{Variance-collapse diagnostics} Table~\ref{tab:variance_diagnostics_ember_az_domain} evaluates variance stability at the block2 layer across incremental domains using the collapse ratio $R$. EMBER-Domain exhibits consistently stable statistics (mean ratio $1.019$) with no significant collapses ($0/11$). AZ-Domain is similarly stable overall (mean ratio $1.078$), with a single significant collapse at Task~1 (ratio $1.3157$) and no collapses thereafter ($1/7$). These results indicate largely preserved feature statistics under interpolation, with instability confined to an early AZ transition.

\section{Broader Impacts}
\label{app:broaderImpacts}

This work contributes to resource-efficient and privacy-conscious CL for evolving malware analysis. By eliminating replay buffers and consolidating knowledge through geometric interpolation, \system reduces the memory and computational overhead typically associated with CL systems while improving long-term detection reliability. These properties are particularly beneficial for practical cyberdefense such as endpoint protection, mobile malware detection, and adaptive threat monitoring, where storage, computation, and data-retention constraints are critical.

As a forgetting-mitigation strategy for malware detection, \system does not introduce any direct negative societal impacts. More broadly, the absence of effective CL mechanisms in evolving threat environments can lead to degraded detection reliability, CF, and reduced adaptability over time, as discussed in Section~\ref{sec:intro}. We hope this work encourages further research into scalable replay-free CL systems for real-world security applications under memory and privacy constraints.

\section{Computational Resources}
\label{app:computation}

All experiments for \system are conducted on a high-performance compute server consisting of NVIDIA RTX A6000 GPUs with 40\,GB VRAM, 7\,TB SATA SSDs, and two NVMe SSDs, running on Ubuntu 20.04.





\end{document}